\def\simlt{\lower.5ex\hbox{$\; \buildrel < \over \sim \;$}}
\def\simgt{\lower.5ex\hbox{$\; \buildrel > \over \sim \;$}}
\RequirePackage{lineno} 
 \documentclass[preprint2]{emulateapj}

\newcommand{\myemail}{mrl@gps.caltech.edu}
\slugcomment{Submitted to the Astrophysical Journal }
\shorttitle{Optimal Estimation}
\shortauthors{Line et al.}
\begin{document}
\title{Information Content of Exoplanetary Transit Spectra:  An Initial Look}
\author{Michael R. Line}
\affil{California Institute of Technology, Pasadena, CA 91125}
\author{Xi Zhang}
\affil{California Institute of Technology, Pasadena, CA 91125}
\author{Gautam Vasisht}
\affil{Jet Propulsion Laboratory, California Institute of Technology, Pasadena, CA 91109}
\author{Vijay Natraj}
\affil{Jet Propulsion Laboratory, California Institute of Technology, Pasadena, CA 91109}
\author{Pin Chen}
\affil{Jet Propulsion Laboratory, California Institute of Technology, Pasadena, CA 91109}
\author{Yuk L. Yung}
\affil{California Institute of Technology, Pasadena, CA 91125}
\email{mrl@gps.caltech.edu}
\altaffiltext{1}{Correspondence to be directed to \myemail}
\begin{abstract}


It has been shown that spectroscopy of transiting extrasolar planets can potentially provide a wealth of information about their atmospheres.
Herein, we set up the {\it inverse} problem in spectroscopic retrieval.  
We use non-linear optimal estimation to retrieve the atmospheric state (pioneered for Earth sounding by Rodgers 1976, 2000). The formulation quantifies the
the degrees-of-freedom and information content of the spectrum with respect to geophysical parameters; herein, we focus
specifically on temperature and composition. First, we apply the technique to synthetic near infrared spectra, and explore the influence of
spectral signal-to-noise ratio and resolution (the two important parameters when designing a future instrument) on the information content of the data.
As expected, we find that the number of retrievable parameters increase with increasing signal-to-noise and resolution, although the
gains quickly level off for large values. Second, we apply the methods to the previously studied atmosphere of HD~189733b, and compare the
results of our retrieval with those obtained by others.

\end{abstract}


\section{Introduction}
Currently there are about 130 confirmed transiting exoplanets (www.exoplanet.org).   Of these planets, several dozen have spectra that have been observed, either through broadband photometry from instruments like the Spitzer Infrared Array Camera (IRAC)  (Knutson et al. 2007; 2008; Harrington et al. 2006; 2007; Stevenson et al. 2011) or higher resolution spectroscopy from the Hubble Space Telescope (HST) Near Infrared Camera and Multi-Object Spectrometer (NICMOS) (Swain et al. 2009a; 2009b), Spitzer Infrared Spectrometer (IRS) (Grillmair et al. 2007; 2008), and recently, from ground based instruments (Swain et al. 2010; Mandel et al. 2011).  Although the spectra are of low resolution ($R= \lambda/\Delta \lambda \sim 5-50$) and low signal to noise (S/N $\le 10$), they nevertheless provide useful information about the temperature and composition of the exoplanetary atmospheres (Tinetti et al. 2007; 2010; Madahusudhan \& Seager 2009; etc.).   A typical approach to retrieving this information is to 
match the data set with forward models by manually tuning the model abundances and temperatures, until a possible best fit is obtained (Tinetti et al. 2007; 2011; Swain et al. 2009a; 2009b).  This approach does not provide an optimal solution to the atmospheric state; furthermore, it can be cumbersome and is susceptible to multiple 
degeneracies (Tinetti et al. 2007; Sing et al. 2008; Madhusudhan \& Seager 2009) 

Others have used multi-dimensional grid models to constrain atmospheric parameters (Madhusudhan \& Seager 2009), a 
method that is well tuned to systematically searching the parameter space given sparse data (as with Spitzer IRAC
color photometry).  In this approach, an ensemble of forward models are generated
using up to 10 gridded free parameters (6 to govern the shape of the temperature profile and 
4 scaling factors for uniform mixing ratios of H$_2$O, CH$_4$, CO, and CO$_2$); model families that best describe the data are selected based on a chi-squared 
statistic criterion.  Because of the degeneracies between the different gases, and between gases and temperature, thousands of solutions can exist within a given
chi-squared region, thus only giving loose constraints on the atmospheric composition and temperature.  Furthermore, the formalism provides no easy way to 
explore the change in information content associated with a change in the data phase space (e.g., $R$ or $S/N$).

Here, we present the {\it inverse} approach (see also Lee et al. 2011) that determines the atmospheric ``state'' (i.e. its temperature structure and abundances) by minimizing 
a cost function that simultaneously takes into account new measurements and prior knowledge of atmospheric properties (such as a 
state retrieved from previous observations).  
Additionally we determine, within the context of our model, the quality of the spectra and the number of useful retrievable atmospheric properties.   
This work represents the first attempt at determining the amount of useful information that can be retrieved from typical exoplanet spectra.  Furthermore, this paper
represents the first attempt at using information theoretic limits for retrievals assuming certain instrument capabilities (such as R and S/N).    Ultimately, the theory is 
general and enables prediction of the advances that can be made with improvements in instrumentation and via more prudent choice of spectral ranges. 

In $\S${2} we outline the basics of the classic retrieval theory of Rodgers (2000).   We first test the technique on an artificial dataset and explore how the number of retrievable parameters depends on $R$ and $S/N$ and discuss how these can be optimized to maximize the usefulness of a measurement in $\S$3.   We then apply these techniques to the well studied HD189733b dayside emission spectra in $\S$4.   This is followed by a discussion and conclusions in $\S$5.


\section{Method}
\subsection{Retrieval Theory}
The retrieval problem is well known in the field of Earth atmospheric studies  (Rodgers 1976, Chahine 1968, Twomey 1977) and in studies of planetary atmospheres (see e.g., Nixon et al. 2007).   The fundamental problem is to determine the state vector, {\bf x} of dimension $n$, often a vector of temperatures and mixing ratios at different altitudes (but could be other desirable variables), given some set of observations, {\bf y} of dimension $m$, usually a vector of flux values at each wavelength.   In the absence of any noise, they can be related through {\bf y=F(x)}, where {\bf F(x)} is a model that simulates the measurement at each wavelength given a representative atmosphere.     In an idealized scenario, if the relationship between  {\bf x} and  {\bf y} is linear, we can linearize {\bf F(x)}  and write
 \begin{equation}
 {\bf y}={\bf F( x_a)}+{\bf K} ({\bf x}-{\bf x_a}) 
 \end{equation} 
where {\bf K}  is the $m \times n$ Jacobian matrix whose elements are given by the Frechet derivative
\begin{equation}
K_{ij}=\frac{\partial F_{i}({\bf x})}{\partial x_j}
\end{equation}
with $F_i$ being the measurement in the $i^{th}$ channel, and $x_j$ the value of the $ j^{th}$ parameter. The vector {\bf x$_a$} is the prior ({\em a priori}) state, which represents our best initial guess of the true state before the observations are made.  The Jacobian describes the sensitivity of the measurement at each wavelength in a spectrum to a perturbation of a given parameter in the forward model.   If the lengths of {\bf x} and {\bf y} are the same then (1) may be readily inverted to
\begin{equation}
 {\bf x}={\bf x_{a}}+{\bf K^{-1}({\bf y} -F({\bf x_a}))}
 \end{equation}
Real data are often noisy and usually have a large number of measurements that over constrain the atmospheric state.  For this we must use a more sophisticated scheme to invert the data to determine the atmospheric properties.  This can be readily achieved by using a Bayesian framework.  In the remainder of this section, we present the basic formalism and useful equations and algorithms that we can use to retrieve atmospheric properties from spectra as well as their information content, following the derivations in Rodgers (2000). For further details, see either Rodgers (2000) or Jacob (2007).  

Bayes theorem can be written as
\begin{equation}
P({\bf x}|{\bf y}) \propto P({\bf y}|{\bf x})P({\bf x})
\end{equation}
where P({\bf x}) is the prior probability distribution, which is knowledge of the atmospheric state before making a measurement, P({\bf y}$|${\bf x}) is the likelihood function, that is the probability that the data exists within the context of a particular model, and P({\bf x}$|${\bf y}) is the posterior probability distribution density function which can be interpreted as the probability that some state {\bf x}, in our case atmospheric state, exists given the observations, {\bf y}.   If we assume Gaussian probability distributions for the observational error and for the {\em a priori} information, we can write 
\begin{equation}
P({\bf y}|{\bf x}) \propto e^{-\frac{1}{2}({\bf y}-{\bf Kx})^{T}{\bf S_{e}^{-1}}({\bf y}-{\bf Kx})}
\end{equation}
\begin{equation}
P({\bf x}) \propto e^{-\frac{1}{2}({\bf x}-{\bf x_{a}})^{T}{\bf S_{a}^{-1}}({\bf x}-{\bf x_{a}})}
\end{equation}
where {\bf S$_{e}$} is the $m \times m$ diagonal error covariance matrix (assuming no correlation between measurements) and  {\bf S$_{a}$} is the $n \times n$ {\em a priori} covariance matrix.  The  {\em a priori} covariance matrix represents our prior knowledge of the natural variability of the system and like {\bf S$_{e}$}, it is assumed to be diagonal.  It essentially defines our ``trust" region, or how far from the prior state we think the actual state can exist.  In general, the prior constraint should be loose enough to allow flexibility in the retrieval but not so loose that the retrieval fails when a measurement contributes no information.

Using Bayes theorem from (4) we can write the posterior probability distribution as a product of (5) and (6)
\begin{equation}
P({\bf x}|{\bf y}) \propto e^{-\frac{1}{2}J({\bf x})}
\end{equation}
where J({\bf x}) is the cost function and is given by
\begin{eqnarray}
J({\bf x})={({\bf y}-{\bf Kx})^{T}{\bf S_{e}^{-1}}({\bf y}-{\bf Kx})} \nonumber \\
+({\bf x}-{\bf x_{a}})^{T}{\bf S_{a}^{-1}}({\bf x}-{\bf x_{a}}) 
\end{eqnarray}
The first term in the cost function represents the contribution from the data.  The second term represents the contribution from the prior knowledge.  If the data is of good quality (high S/N, and high R) then the data term will dominate.   Since the product of two Gaussians is a Gaussian, equation (8) can be equivalently written as
\begin{equation}
J({\bf x})=({\bf x}-{\bf \hat{x}})^{T}{\bf \hat{S}^{-1}}({\bf x}-{\bf \hat{x}})
\end{equation}
where ${\bf \hat{x}}$ and $\bf{\hat{S}}$ are the mean and covariance, respectively, of the posterior probability distribution.  A diagonal element of $\bf{\hat{S}}$ is the variance in the $j^{th}$ component of the  state vector, $\hat{S}_{jj}=\hat{\sigma}^2_j$, where  $\hat{\sigma_j}$ is the retrieval uncertainty in the $j^{th}$ parameter.

The goal of any retrieval is to obtain the most likely set of atmospheric parameters given the data.  This is achieved when (7) is maximized which occurs at the mean of the posterior probability function.   Equating (8) and (9) we can solve for $\bf {\hat{x}}$ and ${\bf \hat{S}}$ to get
\begin{equation}
{\bf \hat{x}}={\bf x_a+ G(y-Kx)}
\end{equation}
where {\bf G} is the gain matrix that describes the sensitivity of the retrieval to the observations (if {\bf G=0}, no sensitivity, then the measurements do not contribute towards the retrieved state) , given by
\begin{equation}
{\bf G}=\frac{\partial {\bf \hat{x}}}{\partial {\bf y}}={\bf \hat{S}K^TS_e^{-1}}
\end{equation}
with 
\begin{equation}
{\bf \hat{S}}={\bf (K^TS_e^{-1}K+S_a^{-1})^{-1}}
\end{equation}
As the elements of ${\bf S_a}$ approach $\infty$ or the elements of ${\bf S_e}$ approach 0, then ${\bf G}$ approaches ${\bf K^{-1}}$ which is identically the sensitivity of the state vector to the observations, and thus the retrieval is fully characterized by the data.   

If the forward model is linear, then (10) can be solved to obtain the desired state vector.  Often, the forward model is non-linear, generally the case in radiative transfer; it is then best to use a numerical iteration scheme to determine the state vector.   In the non-linear case the {\bf Kx} terms in the cost function in (8) are replaced with {\bf F(x)}.  The Levenberg-Marquardt iteration scheme is used to find the minimum of the non-linear cost function.  The prescribed scheme is given by
\begin{eqnarray}
{\bf x_{k+1}}={\bf x_k}+{\bf [}(1+\gamma){\bf S_a^{-1}+K_k^TS_e^{-1}K_k]^{-1}}\nonumber \\
{\bf\{K_k^TS_e^{-1}[y-F(x_k)]-S_a^{-1}[x_k-x_a]  \}      }
\end{eqnarray}
where  ${\bf x_{k}}$ and  ${\bf x_{k+1}}$ are the state vectors for the $k^{th}$ and $k+1^{st}$ iterations, and ${\bf K_k}$ is the Jacobian matrix calculated at the $k^{th}$ iteration.  $\gamma$ is a factor that  controls the rate of convergence and is adjusted at each iteration (Press et al. 1995).  Equation (13) is iterated until convergence, when
\begin{equation} 
{\bf (x_k-x_{k+1})^T\hat{S}^{-1}(x_k-x_{k+1})}<<n
\end{equation}
Upon convergence, we obtain the retrieved state, ${\bf \hat{x}}$ and its precision ${\bf \hat{S}}$.

\subsection{Information Content \& Degrees of Freedom  }
The information content (Shannon \& Weaver 1962) and total number of degrees of freedom are useful quantities that can help diagnose the quality and ability of a spectral data set to contribute to our knowledge of the atmospheric state.  The number of degrees of freedom represents how many independent parameters can be retrieved from the spectrum, and the information content is a metric of how much the precision in the retrieved parameters has improved as a result of the observation.  In the simplest sense, if there are $m$ independent measurements with no error (eg, fluxes at $m$ different wavelengths), then there will be at most be $m$ independent pieces of information (degrees of freedom) that can be obtained from the observations.  If $m$ is fewer than the number of model parameters, $n$, the exact values of $n-m$ parameters cannot be obtained from the observations.  We do not discuss those cases in this article, we choose only cases for which $m > n$.     For a given forward model, with $n$ parameters, the maximum number of obtainable degrees of freedom will be the smaller of $n$ and $m$.   In an ideal case the total number of degrees of freedom will be close to $n$, meaning that the observations can be fully characterized by those $n$ parameters.


In reality, measurements are susceptible error, and the total number of degrees of freedom in the observed signal (denoted by $d_{s}$), and thus the number of parameters accessible to our retrieval, may be fewer than the number of independent measurements, $n$.    Some degrees of freedom,$d_{n}$, can be lost in the noise .  The sum of $d_{s}$ and $d_{n}$ must add up to the total number of parameters we are seeking, $n$.    

Before calculating the degrees of freedom it is useful to first introduce the averaging kernel, {\bf A}.  The averaging kernel tells us which of the parameters in the state vector have the greatest impact on the retrieval, that is, the sensitivity of the retrieval to a given parameter, given by
\begin{equation}
{\bf A}=\frac{\partial {\bf \hat{x}}}{\partial {\bf x}}=\frac{\partial {\bf \hat{x}}}{\partial {\bf y}}\frac{\partial {\bf y}}{\partial {\bf x}}={\bf GK}
\end {equation} 
{\bf A} is an $n\times n$ matrix whose elements are given by 
\begin{equation}
A_{ij}=\frac{\partial \hat{x}_i}{\partial x_j}
\end{equation}
If a diagonal element of {\bf A} is unity, or close to it, then that means for a given change in the true atmospheric state, there is identically the same change in the retrieved state.  This suggests that the parameter, $x_j$, is fully characterized by the data.  If that diagonal element is less than unity, meaning that the data itself is not of a high enough quality to constrain that parameter, then some fraction of the {\em a priori} information must have been used in determining the value of that parameter.  If each parameter is fully characterized by the data, that is if, all of the diagonal elements of {\bf A} are unity, then we would expect to be able to retrieve all $n$ parameters.  If the diagonal elements are less than unity, then the sum of the diagonals would be less than $n$.   In essence, the diagonal elements of the averaging kernel can be thought of as the degrees of freedom per parameter.  If the value of a particular diagonal element is 1, then that parameter is well characterized by the data.  If it is much less than 1, then the data contributes little to our knowledge of that parameter.    The total degrees of freedom from the signal can be determined by calculating the trace of {\bf A}.  The difference between $n$ and the trace of {\bf A} is the number of degrees of freedom lost to the noise.  

The total degrees of freedom, again, tell us how many independent parameters can be determined from the observations.  The information content, $H$, tells us quantitatively how well the observations increased our confidence in our estimate of the atmospheric state relative to the {\em a priori} knowledge.   In a more precise language, the information content of a measurement is the reduction in the entropy of the probability that that an atmospheric state exists given some set of observations, or
\begin{equation}
H=entropy(P({\bf x}))-entropy(P({\bf x|y}))
\end{equation}
The entropy of a Gaussian distribution of width $\sigma$,  which the prior and a posterior distributions are assumed to be, can be shown to be proportional to $ln(\sigma)$.  Using this fact, and equations (17), (6), and (9),
\begin{equation}
H=\frac{1}{2}ln({\bf |\hat{S}^{-1}S_a}|)
\end{equation}
From this we can see that if the data is good (small error bars), then the elements of ${\bf \hat{S}}$ will be small, resulting in a large $H$.    Thus $H$ is a quantitative measure of the reduction in our uncertainty in the retrieved atmospheric state as a result of the observations.   The larger the value of $H$, the more useful the observations are in constraining the atmospheric state.  

In summary, both $d_s$ and $H$ are quantitative measures of the quality and usefulness of the observations in determining the atmospheric state, within the context of a given forward model.  From their definitions we would expect that a spectrum with a higher S/N, or a higher R, would result in higher values.  We will show this in section $\S$3.


\subsection{Forward Model}
A relatively simple forward model, ${\bf F(x)}$, which nonetheless captures the basic physics and
the measurement process, is at the core of our retrieval. We assume a simplified understanding
of the physical and chemical state of the exoplanet atmosphere, i.e., a parameterized temperature
structure, the major volatile constituents, the important radiative processes, and the instrument 
line profiles etc.  Our forward model, as most such models, is an approximation because the data are of limited quality, the underlying
physics is relatively ill-understood, and simplifying approximations are necessary.
Examples of physics missing in our ${\bf F(x)}$ include absent species, inaccurate line lists, clouds, aerosols, 3D effects etc., or 
possibly insufficient parameterization of the atmosphere.   Therefore, our retrievals must be taken in context of our chosen forward model.  
Herein, we only consider the dayside spectra of hot-Jupiters with near solar metallicity, though the methods are easily be extended to
other kinds of observations (transmission spectra) and exoplanets (hot-Neptunes, mini-Neptunes, super-Earths etc.) with 
relatively minor modifications to the forward model. For future instruments, with broader spectral coverage and
higher spectral resolution, the forward models can increase in sophistication.

Lacking sufficient data (these are low signal-to-noise, low resolution spectra), we simplify our atmosphere to 8 parameters that characterize the temperature structure and gas concentrations.  
For sake of simplicity, we use an analytic temperature profile formulated by Guillot (2010), and since then modified by Parmentier \& Guillot, (in preparation) 
to include three channels. The profile, derived using a 3 channel approximation, is given by
\begin{equation}
T^4(\tau)=\frac{3T^4_{int}}{4}(\frac{2}{3}+\tau)+\frac{3T^4_{irr}}{4}(1-\alpha)\xi_{\gamma_{1}}(\tau)+\frac{3T^4_{irr}}{4}\alpha \xi_{\gamma_{2}}(\tau)
 \end{equation}
where
\begin{equation}
\xi_{\gamma_{i}}=\frac{2}{3}+\frac{2}{3\gamma_{i}}[1+(\frac{\gamma_{i}\tau}{2}-1)e^{-\gamma_{i}\tau}]+\frac{2\gamma_{i}}{3}(1-\frac{\tau^{2}}{2}){\rm E_{2}}(\gamma_{i}\tau)
\end{equation}
with $\gamma_{1}=\kappa_{v_{1}}/\kappa_{IR}$ and  $\gamma_{2}=\kappa_{v_{2}}/\kappa_{IR}$, where $\kappa_{v_{1}}$, $\kappa_{v_{2}}$, and $\kappa_{IR}$ are the visible and infrared (thermal) opacities, respectively.  The parameter $\alpha$ (range 1 to 0) partitions the flux between the two visible streams, and
$E_2(\gamma \tau)$ is the second order exponential integral function.  The internal heat flux (from the net cooling history) is represented by the temperature T$_{int}$, while the solar flux at the top of the atmosphere is represented by T$_{irr}$;  these two temperatures are fixed.  Assuming zero albedo and unit emissivity, $T_{irr}$ is 
\begin{equation}
T_{irr}=(\frac{R_{*}}{2a})^{1/2}T_{*}
\end{equation}
where $R_{*}$ and $T_{*}$ are the stellar radius and temperature, $a$, the star planet separation and $\tau$ is the infrared (thermal) optical depth 
\begin{equation}
\tau=\frac{\kappa_{IR}P}{g}
\end{equation}
with $P$ the pressure and $g$ the surface gravity (at 1 bar).   In total there are 4 free parameters governing the temperature structure,
$\kappa_{IR}$, $\kappa_{v_{1}}$, $\kappa_{v_{2}}$ and $\alpha$.   We choose this parameterization with two visible streams as opposed to the traditional one visible stream (Hansen 2005; Guillot 2010) because the extra stream allows more freedom for a temperature inversion, though in some cases (as we shall see below) the second visible stream does not matter.

The remaining 4 parameters are the uniform mixing ratios for H$_2$O, CH$_4$, CO, CO$_2$, expected to be the major molecular opacity sources (Tinetti et al., 2007; Swain et al., 2009a).   We choose vertically uniform mixing ratios for two reasons.  First, the data lack sufficient information content to actually help resolve vertical structure in abundances, and second, chemical kinetics models (Moses et al. 2011; Line et al. 2010, 2011), show that vertical mixing leads to constant vertical mixing ratios for these species within the IR photosphere, so even if we could resolve detailed vertical information, we would most likely find that the abundances remain fairly constant.   

Since many of these parameters may vary over many orders of magnitude we find it convenient with the above formalism to solve for the logarithm 
of the atmospheric state.  With that, the state vector of parameters that we would like to retrieve can be given by
\[ {\bf x} = \left[ \begin{array}{ccc}
\log (\kappa_{v_{1}}) \\
\log (\kappa_{v_{1}}) \\
\log (\kappa_{IR}) \\
 \alpha \\
\log ( f_{H2O})\\
\log ( f_{CH4})\\
\log ( f_{CO})\\
\log ( f_{CO2})\\
  \end{array} \right]\] 
where $f_{i}$ is the mixing ratio of species $i$ in parts per million (ppm) and the opacities are in cm$^2$g$^{-1}$. 

We also include H$_2$-H$_2$ and H$_2$-He collision induced opacity. The mixing ratios of H$_2$ and He vary little with the atmospheric levels that 
produce the bulk of the dayside thermal emission  
(500-2000 K, 10-10$^{-4}$ bar). We fix $f_{H_2}$ and $f_{He}$ to thermochemical abundances (assuming solar elemental abundances) of  0.86 and 0.14, respectively.  
These values may change on the tens of percent level in enriched atmospheres, however, this variation has negligible effect on the
resultant infrared spectra. Also, we do not include NH$_3$ as an opacity source as it has little influence in the spectral region we consider.    

We use the Reference Forward Model (RFM)\footnote[2]{see http://www.atm.ox.ac.uk/RFM/}, a line-by-line radiative transfer code, to calculate the disk integrated dayside emission spectra, modified to handle H$_2$-H$_2$ and H$_2$-He collisionally induced opacities.  The collisionally induced opacity tables are taken from Barysow et al. (2001;2002) and J{\o}rgensen et al. (2000).   The molecular line strengths for H$_2$O, CO$_2$, and CO, are from the HITEMP (Rothman et al. 2010) database and CH$_4$\footnote[3]{Upon completion of our initial investigation it was also brought to light that there exists more appropriate high temperature based line lists for methane such as the STDS (http://icb.u-bourgogne.fr/OMR/SMA/SHTDS/HTDS.html).  Using this line list over HITRAN makes absolutely no difference for our synthetic work since the synthetic data was produced using the HITRAN methane.  We have also compared our HD189733b retrieval results for both methane line lists and found no difference.} is from the HITRAN 2008 database (Rothman et al. 2009).   In order to keep the molecular line-lists from becoming too unwieldy we make an intensity cutoff at 298 K of 10$^{-40}$  cm molecule$^{-1}$,  as recommended by Sharp \& Burrows (2007).

\section{Test on Synthetic Data}

First, we test the retrieval method on a synthetic data set for which we know the answer.  Using this synthetic spectrum,  
 we explore the effect that signal-to-noise and spectral resolution have on the degrees of freedom and information content.  

A hypothetical hot-jupiter atmosphere is generated using $\kappa_{v_{1}}=\kappa_{v_{2}}=4\times10^{-3}$ cm$^2$g$^{-1}$, $\kappa_{IR}=1\times10^{-2}$ cm$^2$g$^{-1}$, $\alpha=0.5$, and fixed vertical mixing ratios of $f_{H2O}=5\times10^{-4}$, $f_{CH4}=1\times10^{-6}$, $f_{CO}=3\times10^{-4}$, and $f_{CO_2}=1\times10^{-7}$. 
The planet orbits around a G0V host star (e.g. HD~209458a) with $T_{*}=6000$ K, $R_{*}=1.14$ R$_\odot$ at a separation of $a = 0.064$ AU.  
The planetary properties are a radius of 1.35$R_{J}$, an internal temperature of $T_{int}=200$ K, and $g = 21.1$ m s$^{-2}$ (at 1 bar pressure).  
Using (21) we find $T_{irr}=1223$ K.   The emission spectrum of the exoplanet (see Figure 1) is initially generated 
with a one wave-number resolution (resolving power,  R $\simeq $5000 at 2 $\mu m$).

\begin{figure}
  \centering
    \includegraphics[width=0.5\textwidth]{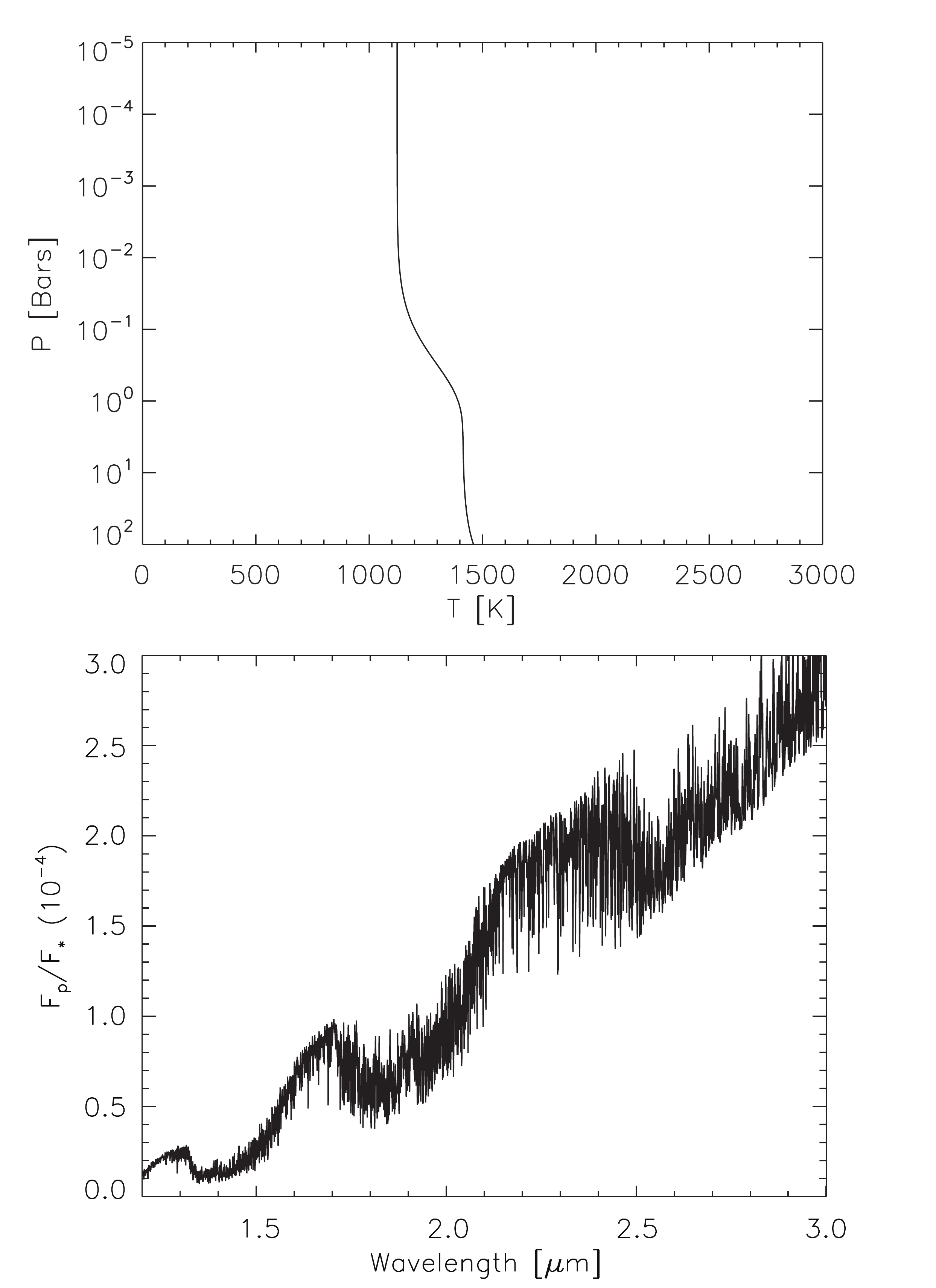}
     \caption{Synthetic spectrum (bottom) generated with the model atmosphere (top) with a spectral resolution of 1 cm$^{-1}$, or R$\sim$5000 at 2 $\mu$m.  The model temperature profile is generated from equations (19) and (20) with $\kappa_{v_{1}}=\kappa_{v_{2}}=4\times10^{-3}$ cm$^2$g$^{-1}$, $\kappa_{IR}=1\times10^{-2}$ cm$^2$g$^{-1}$, $\alpha=0.5$, $T_{irr}=1223K$, and $T_{int}=200K$.  The constant-with-altitude mixing ratios are $f_{H2O}=5\times10^{-4}$, $f_{CH4}=1\times10^{-6}$, $f_{CO}=3\times10^{-4}$, and $f_{CO2}=1\times10^{-7}$.   }
\end{figure}

For the initial test, the synthetic spectrum (Figure 1) is degraded by convolving it with an instrumental profile 
matching the defocussed HST NIC3 camera with a spectral full width at half maximum
of 0.055 $\mu$m (R $\simeq$ 40 at 2 $\mu$m; Swain et al. 2009a), and reducing the measurement
signal-to-noise of each spectral channel to $\sim 10$. Rather than be guided by physical and chemical
models, or some previous observation of the object, 
we arbitrarily chose an {\em a priori} state, ${\bf x_a}$, far from the true physical state.  The remaining 
unspecified quantity is the {\em a priori} covariance matrix, ${\bf S_a}$. Once more, the diagonal
elements of ${\bf S_a}$ are allowed a large range as we are dealing with a relatively novel type
of observations and lack detailed prior information. We also assume that there are no cross correlations 
between different state parameters (e.g. $f_{CO}$ and $f_{CO_2}$, even though from chemical models  we
know that such quantities have high correlations).   
Because the state parameters are logarithmic, the elements of ${\bf S_a}$ are also logarithmic  (with the exception of $\alpha$) so we set, somewhat arbitrarily, $\sigma_{\kappa_{v1}}=2$,  $\sigma_{\kappa_{v2}}=2$ ,  $\sigma_{\kappa_{IR}}=2$ ,  $\sigma_{\alpha}=0.5$ ,  $\sigma_{f_{H2O}}=6$ ,  $\sigma_{f_{CH4}}=6$ , $\sigma_{f_{CO}}=6$, and $\sigma_{f_{CO2}}=6$ meaning that the opacities are permitted to span 4 orders of magnitude centered around their {\em a priori} value and the mixing ratios are allowed to span 12 orders of magnitude.    Such large {\em a priori} uncertainties lead to a flat {\em a priori} distribution, relative to the data,
reducing the current problem to a maximum likelihood estimation (as opposed to Bayesian), with the option of using the priori information if the data is sparse.

The entirety of the forward model can summarized with the Jacobian.  Figure 2 shows the columns of the Jacobian evaluated at the true state (response of the flux in each channel to a perturbation in each of the parameters in {\bf x}) for the synthetic data (Figure 3).   The spectrum is most sensitive to perturbations in the opacities that govern the temperature profile.  The 1.7 $\mu$m and 2.2 $\mu$m channels are most sensitive to changes in the temperature profile.  This is because there aren't large absorption features at these wavelengths, meaning, these channels are most sensitive to the flux from deeper layers (1-10 bars).  This also partially explains why $\kappa_{IR}$ and $\kappa_{v1}$ have opposite responses.  An increase in $\kappa_{IR}$ results in an increase in flux due to an increase in temperature in the deep layers probed by these channels, as can be seen in (19).  An increase in $\kappa_{v1}$ results in a decrease of flux in these channels due to a decrease in temperature in the deeper layers.  From (19) an increase in $\kappa_{v1}$ increases the temperature above the $\sim 0.1$bar level, and in order to maintain radiative equilibrium at the top of the atmosphere, a decrease in temperature in the deeper layers must occur, and also a higher $\kappa_{v}$ prevents the stellar flux from penetrating into the deeper atmosphere.  The opposite is true near 2.9 $\mu$m which is more sensitive to higher altitudes because of the large absorption, thus an increase in in $\kappa_{v1}$ will result in an increase in temperature which in turn results in a flux increase.  Also, in this particular case $\alpha=0.5$ meaning both $\kappa_{v1}$ and $\kappa_{v2}$ have identically the same results.  Additionally, $\kappa_{v1}$=$\kappa_{v2}$ which causes the spectrum to have no sensitivity to changes in $\alpha$. 

The spectral response is most sensitive to the water abundance more than any other gas across all wavelengths in this example (Figure 2).  This makes the retrieval of water more precise than the other species.  The greatest sensitivity to changes in the CO$_2$ abundance occur at 2.1 and 2.8 $\mu$m, which both happen to be located near the sensitivity minima of CO and CH$_4$, though it still has to contend with water.  Both CO and CH$_4$ have greatest sensitivity in the 2.3 $\mu$m band making it difficult to simultaneously retrieve both.

\begin{figure}
  \centering
    \includegraphics[width=0.5\textwidth]{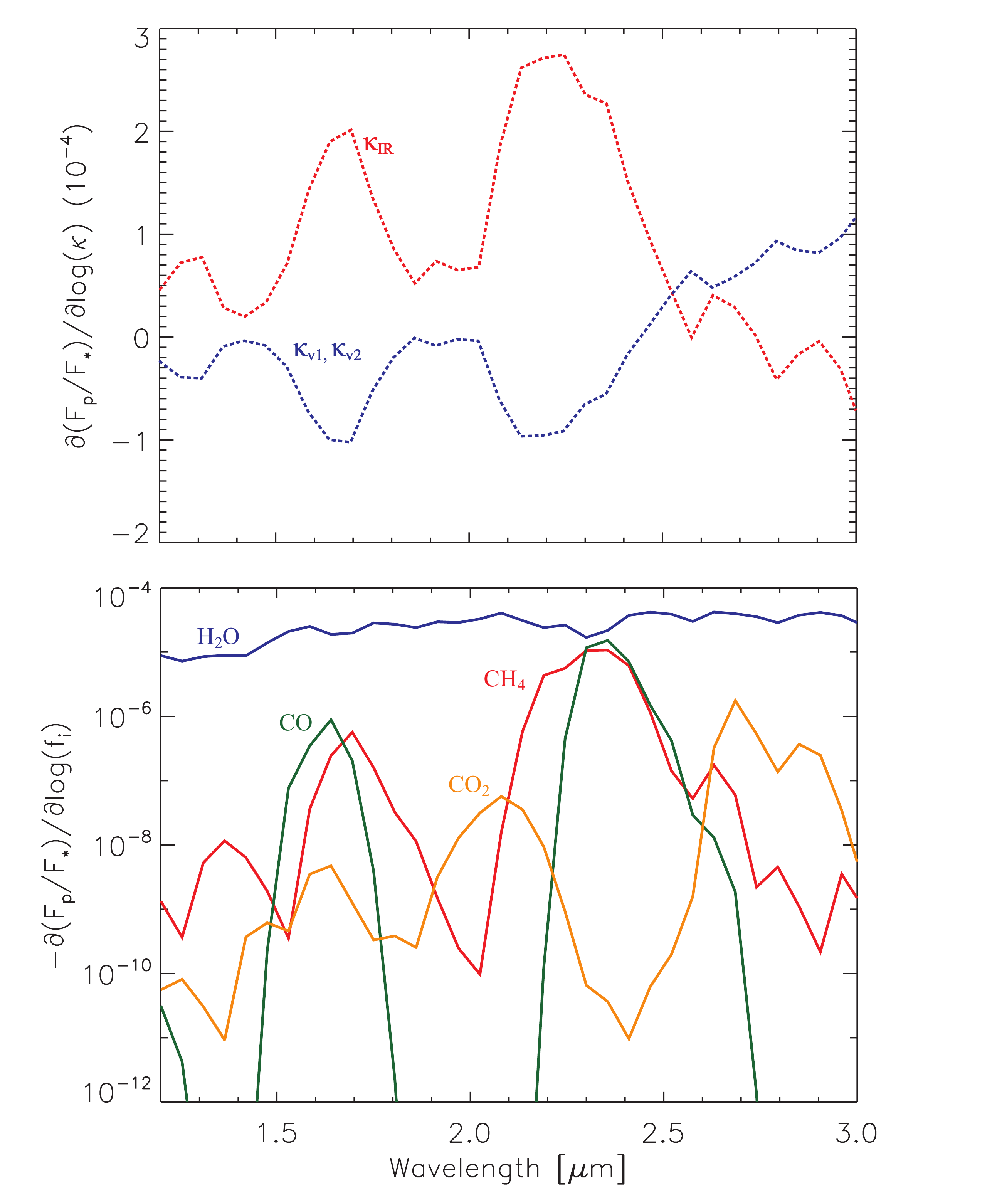}
     \caption{Columns of the Jacobian for the synthetic spectrum evaluated at the true state.  This is the response of the flux as a function of wavelength due to a small positive perturbation in one of the parameters in {\bf x}.  The top panel is the flux response for the parameters that govern the temperature profile, $\kappa_{v_{1}}$, $\kappa_{v_{2}}$, $\kappa_{IR}$.  The bottom panel is the flux response to a small perturbation in the gas mixing ratios,  $f_{H2O}$, $f_{CH4}$, $f_{CO}$, and $f_{CO2}$.  The Jacboian is calculated as a change in the planet-to-star flux ratio, $\Delta (F_p/F_{*})$ to a positive logarithmic perturbation in a given parameter, $\Delta \log(x_j)$.  Note that in the bottom panel an increase in the gas mixing ratios always results in a decrease in $F_p/F_{*}$.  In this particular case, the spectrum is equally sensitive to $\kappa_{v_{2}}$ and $\kappa_{v_{1}}$ because $\alpha$ is 0.5.  If $\alpha=0$ than the spectrum will have no sensitivity to $\kappa_{v_{2}}$ and if $\alpha=1$ the spectrum will have no sensitivity to  $\kappa_{v_{1}}$.  Also, for this synthetic dataset $\kappa_{v_{2}}$=$\kappa_{v_{1}}$ which results in no sensitivity to $\alpha$   }
\end{figure}

Figure 3 shows the retrieval process for this initial synthetic test case.   We determine the quality of the retrieval using the standard reduced chi-squared given by
\begin{equation}
\chi^2=\frac{1}{N}\sum_{i=1}^N \frac{(y_i-F_i)^2}{\sigma_i^2}
\end{equation}
where $N$ is the total number of data points, $y_i$, $F_i$, and $\sigma_i$, are defined in $\S$2.1.  If $\chi^2$ is less than one, then the difference between the model fit and data is typically better than 1 $\sigma$.   We should stress however, that a perfect fit ($\chi^2=0$) does not necessarily mean that the true state has been retrieved, because of the degeneracies between some of the parameters.  Table 1 compares the true state to the retrieval results along with the retrieval precission.   The synthetic retrieval demonstrates the robustness of the retrieval to a poor {\em a priori}.   The reason for this can be seen by inspecting the elements of the averaging kernal.  From Table 1, all but $\kappa_{v1}$ and methane are fairly well characterized by the data ($A_{jj}$ is close to 1).  Summing these values gives the total degrees of freedom, and thus the total number of useful retrievable parameters of $\sim 6$.

\begin{figure*}
  \centering
    \includegraphics[width=1\textwidth]{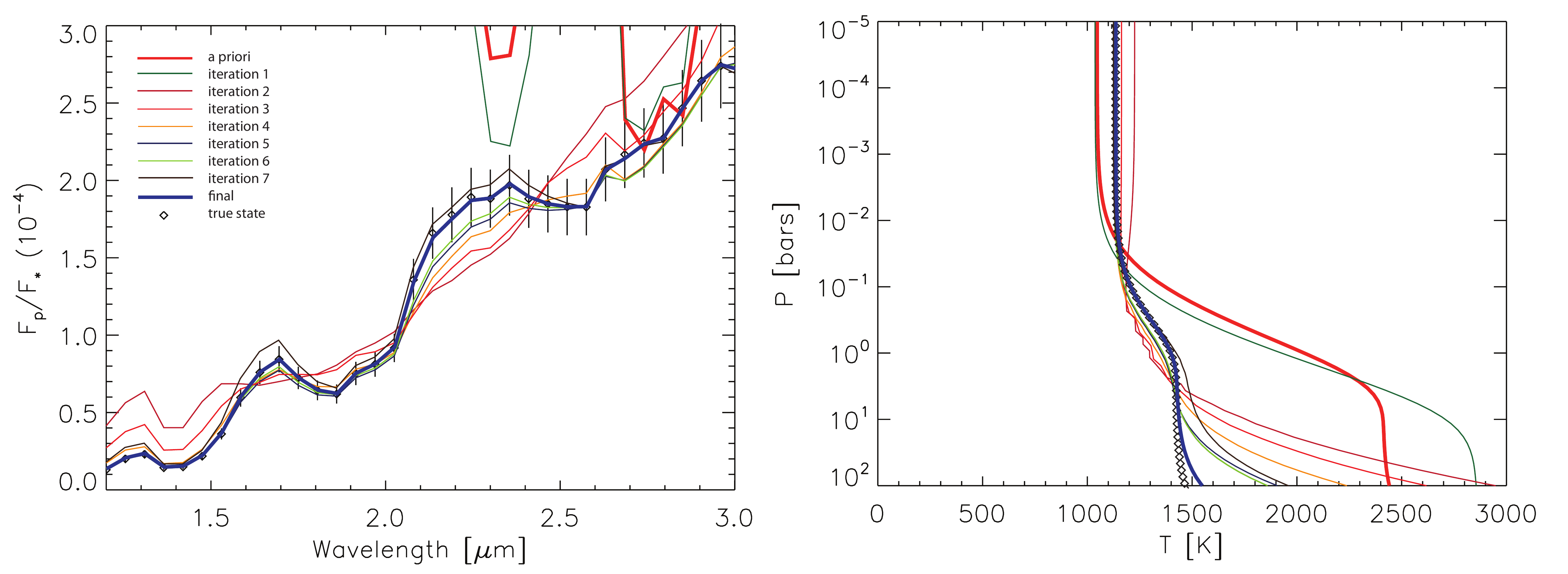}
     \caption{Synthetic spectrum retrieval.  Left: Iteration sequence of the model spectrum, ${\bf F(x_k)}$.  The diamonds with error bars are the synthetic data convolved down to a resolution of 0.055 $\mu$m (R$\sim$37 at 2 $\mu$m) and a signal-to-noise of 10.  The thick red curve is the forward model spectrum generated from the {\em a priori}, ${\bf F(x_a)}$.  Note that it is a poor fit to the data.  Each subsequent curve is the new model spectrum after each iteration of equation 13.  The thick solid blue curve is the final retrieved model spectrum. Right:  Evolution of the temperature profile with each iteration.  The thick red curve is the {\em a priori} temperature profile.  The thick blue curve is the retrieved temperature profile.  The diamond symbol curve is the true temperature profile as in figure 1.   $\chi^2$ converges to 0.007 after 8 iterations of equation 13.   }
\end{figure*}

\begin{center}
\begin{deluxetable*}{ccccccc}
\tablecolumns{6} 
\tablewidth{0pt}  
\tablecaption{Synthetic retrieval results.  $\kappa_{v1}$, $\kappa_{v2}$, and $\kappa_{IR}$ are in units of (cm$^2$g$^{-1}$).  $f_{i}$ is the volume mixing ratio for species $i$.  We also show the diagonal averaging kernal elements ($A_{jj}=\frac{\partial \hat{x}_j}{\partial x_j}$) for each parameter.  The retrieval uncertainties are given as $\hat{x}-\hat{\sigma}$ to $\hat{x}+\hat{\sigma}$ for each parameter.
      }
\tablehead{\colhead{Parameter }			&
		\colhead {True State ({\bf x})  }                                    &
				\colhead{{\em A priori} (${\bf x_a}$)}		&
                 \colhead{Retrieved State (${\bf \hat{x}}$)}                             &
                \colhead{Retrieval Precision }                   		&
                     \colhead{$\frac{\partial \hat{x}_i}{\partial x_j} $}                   &
                           }
\startdata
$\kappa_{v1}$& 4.00$\times$10$^{-3}$&1.00$\times$10$^{-3}$  & 3.59$\times$10$^{-3}$  &2.76$\times$10$^{-3}$ - 4.68$\times$10$^{-3}$ &0.997 &\\
$\kappa_{v2}$ & 4.00$\times$10$^{-3}$ &1.00$\times$10$^{-2}$  & 1.70$\times$10$^{-9}$  & 1.70$\times$10$^{-11}$ - 1.70$\times$10$^{-7}$&0.0\\
$\kappa_{IR}$  & 1.00$\times$10$^{-2}$ & 3.16$\times$10$^{-2}$ &8.93$\times$10$^{-3}$ & 7.13$\times$10$^{-3}$ - 1.12$\times$10$^{-2}$&0.998\\
$\alpha$  &0.5 &  0.1 & 0.003 &0.00 - 0.022 &0.999\\
$f_{H2O}$  & 5.00$\times$10$^{-4}$&1.00$\times$10$^{-6}$  & 4.18$\times$10$^{-4}$&2.58$\times$10$^{-4}$ - 6.76$\times$10$^{-4}$ &0.999\\
$f_{CH4}$  &  1.00$\times$10$^{-6}$&1.00$\times$10$^{-4}$  &3.43$\times$10$^{-7}$ & 4.34$\times$10$^{-12}$ - 2.70$\times$10$^{-2}$&0.334\\
$f_{CO}$  & 3.00$\times$10$^{-4}$ & 1.00$\times$10$^{-6}$ &1.96$\times$10$^{-4}$ &2.27$\times$10$^{-6}$ - 1.69$\times$10$^{-2}$ &0.896\\
$f_{CO2}$  &  1.00$\times$10$^{-7}$&1.00$\times$10$^{-4}$  & 7.70$\times$10$^{-7}$&9.95$\times$10$^{-10}$ - 5.96$\times$10$^{-4}$ &0.768\\
\enddata
\end{deluxetable*}
\end{center}

\subsection{Resolution and Signal to Noise Effects on the Degrees of Freedom \& Information Content}
The S/N and R are two important factors that influence the quality and usefulness of a spectrum.  It is thus imperative to consider them when designing a spectrometer.  In this section we use our synthetic dataset to explore how the degrees of freedom, both total and per atmospheric parameter, and the information content evolve with increasing S/N and R.   

We would intuitively expect  $d_s$ and $H$ both to increase with increasing R and S/N.  Figure 4 shows a contour plot of $d_s$ and $H$ calculated for the synthetic spectrum generated in Figure 1 for a variety of S/N's and R's.  The maximum increase in both occurs with a simultaneous\footnote[4]{This is true if R and S/N are independent of each other.  In most cases S/N decreases with increasing R because of the smaller spectral bins. } increase in S/N and R.   

We point out that the contour plots in Figure 4 can only be taken in the context of the spectral window within which we are applying the retrieval, and the number of parameters we are trying to retrieve.   In other words, for the 8 parameters we are retrieving here, there is no benefit to increasing R or S/N beyond a few hundred and $\sim$100, respectively.  If we do happen to have a higher R and S/N, it is likely that we would be able to retrieve more forward model parameters such as the concentrations of other gases, or information on the vertical distributions of the gases.   Current observations, like the HST NICMOS observations of HD189733b, generally fall towards the bottom left corners in Figure 6.  This suggests that S/N and R's of such data are not high enough to fully constrain even our simple forward model, and thus even less constraining for more complicated models.      

\begin{figure*}
  \centering
    \includegraphics[width=1\textwidth]{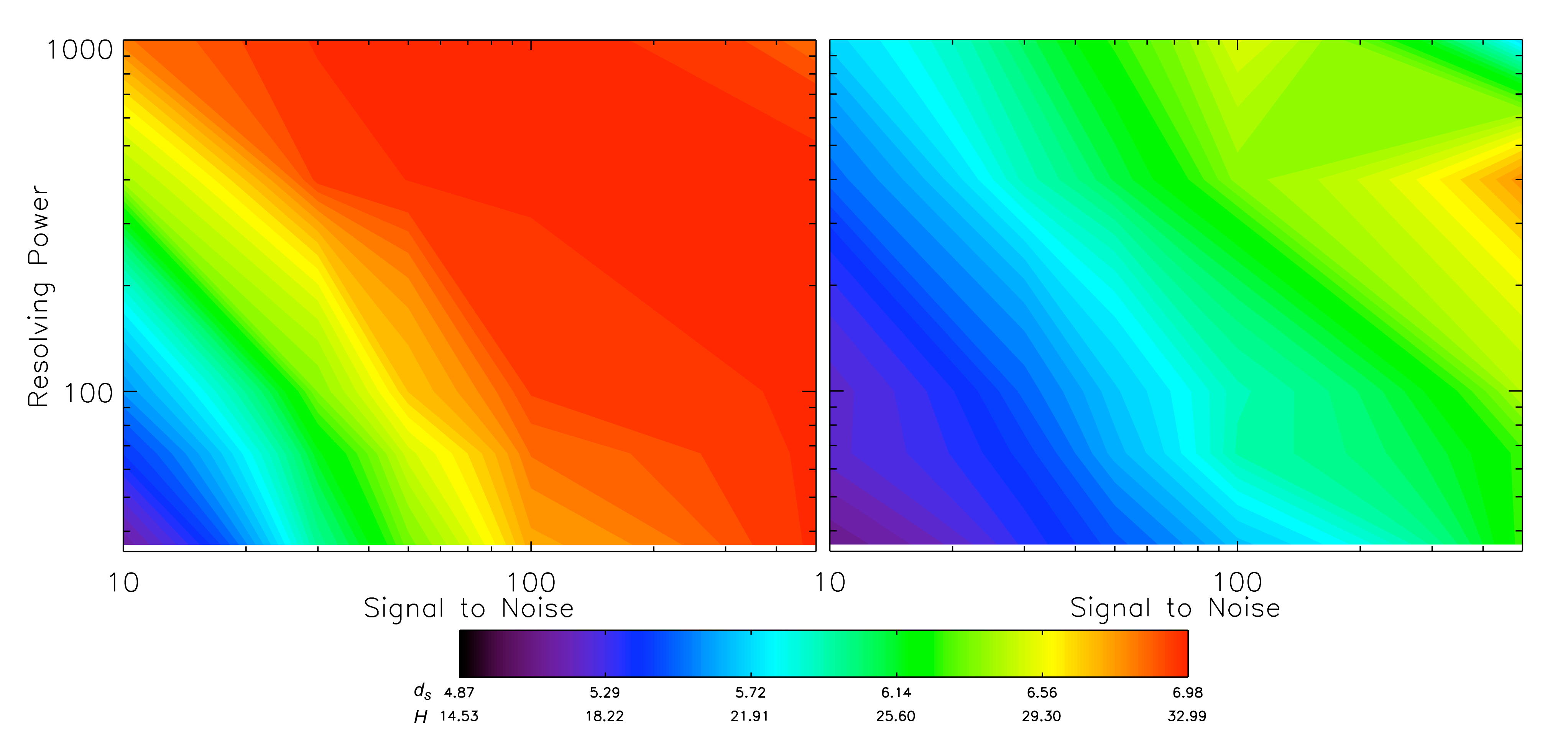}
     \caption{S/N and R effects on the total degrees of freedom (left) and the information content (right).  In general, as S/N and R increase, the total number of degrees of freedom obtainable from the data, and the information content increase.  See equations (24) and (25).  }
\end{figure*}

The increasing behavior in $d_s$ with increasing S/N can be seen through the use of (11), (12), and (15).  As S/N goes to infinity, the elements of ${\bf S_e}$ go to zero causing {\bf G} to approach ${\bf K^{-1}}$, in turn causing {\bf A} to approach the identity matrix, meaning the diagonal elements are all ones with a trace equal to the total number of parameters and thus the maximum number of degrees of freedom.   The relationship between $d_s$ and S/N can be seen in a 1-parameter 1-channel model, where $d_s=A$.  Upon reducing the matrix equations, the one element averaging kernel becomes, 
\begin{equation}
d_s=A=\frac{K^2\sigma_a^2}{K^2\sigma_a^2+(F/(S/N))^2}=\frac{(S/N)^2}{(S/N)^2+\frac{F^2}{K^2\sigma_a^2}}
\end{equation}
and the relation of these parameters to the information content is
\begin{equation}
H=ln[1+\frac{\sigma_a^2}{F^2}K^2(S/N)^2].
\end{equation}
where K, $\sigma_a$, and F are the 1-D analogs for {\bf K}, ${\bf S_a}$, and {\bf F(x)}, respectively.  We also have assumed that $\sigma_e$, the 1-D analog for ${\bf S_e}$, is the flux, $F$, divided by S/N.  In this case, $d_s$ approaches unity as S/N goes to infinity, and zero, if S/N is zero.  $H$ approaches infinity as S/N goes to infinity, and approaches zero when S/N goes to zero.  One important thing to note from these relations is that increasing S/N will matter only if the Jacobian, K, is non-zero, meaning that there must be some sensitivity of the flux to a perturbation in the desired parameter.  Otherwise, no amount of S/N increase will improve our knowledge of the atmospheric state.   
Increasing R or adding more spectral channels can also contribute to an increase in $d_s$ and H.  If channels are chosen such that the K is large, meaning large sensitivity to a given parameter, then $d_s$ and $H$ will both increase.  As K approaches infinity (infinite sensitivity), $d_s$ will approach unity and $H$ will approach infinity. 

From this simple analysis, though it may intuitively obvious, we can readily see that if we want to improve the characterization of a particular atmospheric property, it is best to design an instrument whose spectral regions offer the greatest sensitivity to that parameter, and to have a high S/N within those spectral regions.  

\section{Test on Real Data: HD189733b Dayside Emission}
Now that we have demonstrated that this retrieval procedure works and provides useful information about the quality of a data set through the degrees of freedom and information content, we wish to apply it to the dayside emission spectra of one of the best-studied exoplanet atmospheres, HD189733b.   We assume the same forward model and {\em a priori} covariances as in the synthetic work.   

The dayside emission spectrum of HD189733b has been subject to much investigation (Swain et al., 2009a, Grillmair et al. 2007,  Madhusudhan \& Seager 2009, and many others), and often times different analyses come up with different solutions for its composition and temperature structure.   For simplicity we investigate only the near IR spectrum from Swain et al. (2009a).   As an {\em a priori} atmospheric state we use the ``Fortney 2$\pi$"  (Fortney et al., 2010) temperature profile from Figure 2 of Moses et al. (2011) approximated with equation (19) and the 0.1 bar mixing ratios for H$_2$O, CH$_4$, CO, and CO$_2$ from their table 2 but assumed to be constant with altitude within the IR photosphere sampled by the observations (because of quenching arguments).   Figure 5 and Table 2 show the results of the retrieval.  The Jacobian in Figure 5 demonstrates the high sensitivity of the spectrum to water and carbon dioxide, some sensitivity to CO near 2.3 $\mu$m, and very little sensitivity to methane at all wavelengths.  The 1.7 and 2.2 $\mu$m channels are sensitive to the deep temperatures (effected by $\kappa_{IR}$) due to the higher transmittance at those wavelengths.  The strong CO$_2$ absorption feature at 2.1 $\mu$m has less sensitivity to the deep temperatures and more sensitivity to temperatures higher up (controlled by $\kappa_{v1}$ and $\kappa_{v2}$).  

The diagonal elements of the averaging kernel in Table 2 quantitatively tell us which parameters we can and cannot retrieve from the dayside emission spectra.  Again, H$_2$O, CO and CO$_2$ have averaging kernel elements that are near unity and are therefore well constrained by the data, as is also reflected in the retrieval uncertainty, which is smaller than the assumed {\em a priori} uncertainty.    CH$_4$ is completely unconstrained.  The retrieval uncertainty is the same as the {\em a priori} uncertainty, suggesting that the observations contribute no information about its abundance.  The trace of the averaging kernel gives the total number of degrees of freedom, and thus the total number of retrievable parameters, to be  $\sim$5.  

Our results compare quite well with those of Madhusudhan \& Seager (2009) and with Swain et al. (2009a) with the exception of CO$_2$ (Table 2) which appears to be underestimated by three orders of magnitude in Swain et al. (2009a).   Our derived temperature profile (Figure 5, bottom right) also appears to fall within the spread given in Figure 5 of Madhusudhan \& Seager (2009).

\begin{figure*}
  \centering
    \includegraphics[width=1\textwidth]{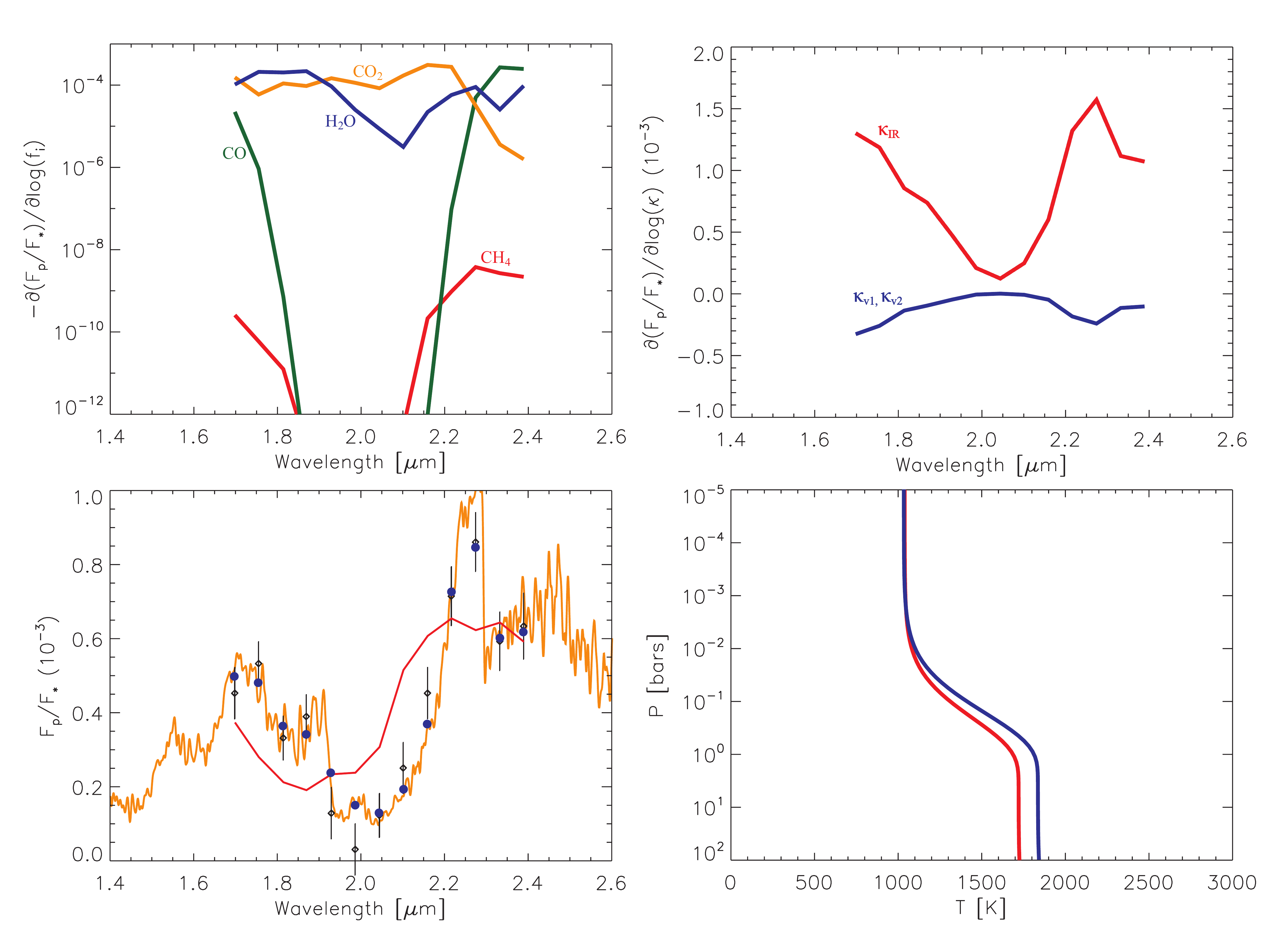}
     \caption{Retrieval results for the NICMOS dayside emission spectra of HD189733b from Swain et al. (2009a).   Top Left:  The sensitivity of the planet-to-star flux ratio to a perturbation in the mixing ratios of H$_2$O, CO$_2$, CO, and CH$_4$ at each channel in the NICMOS dataset.  Top Right:  The sensitivity of the planet-to-star flux ratio to a perturbation in the parameters governing the temperature profile.  Bottom Left: The retrieved spectrum.  The black diamonds with error bars are the Swain et al. (2009a) dayside emission data. The red curve is the {\em a priori} spectrum convolved with the instrumental broadening profile and sampled at the  data wavelengths.  The orange curve is retrieved spectrum at high resolution.  The blue dots are the retrieved spectrum convolved with the instrumental broadening function and sampled at the data wavelengths.  This optimal solution gives  $\chi^2$=0.76.  Bottom Right:  The {\em a priori} (red) and retrieved (blue) temperature profiles.  }
\end{figure*}

\begin{center}
\begin{deluxetable*}{ccccccccc}
\tablecolumns{5} 
\tablewidth{0pt}  
\tablecaption{Retrieval results for HD189733b.  $\kappa_{v1}$, $\kappa_{v2}$, and $\kappa_{IR}$ are in units of (cm$^2$g$^{-1}$).  $f_{i}$ is the volume mixing ratio for species $i$.  We also show the diagonal averaging kernal elements ($A_{jj}=\frac{\partial \hat{x}_j}{\partial x_j}$) for each parameter.   The total number of degrees of freedom for this spectrum is $\sim$5.  The retrieval precisions are given as $\hat{x}-\hat{\sigma}$ to $\hat{x}+\hat{\sigma}$ for each parameter.  We also show for comparison the abundances derived by Madhusudhan \& Seager (2009) (MS10) and Swain et al. (2009a) (S09a).
      }
\tablehead{\colhead{Parameter }			&
		\colhead { {\em A Priori} ({\bf x$_a$})  }                                    &
                 \colhead{Retrieved State (${\bf \hat{x}}$)}                             &
                \colhead{Retrieval Precision }                   		&
                     \colhead{$\frac{\partial \hat{x}_i}{\partial x_j} $}                   &
                    \colhead{MS10} &
                   \colhead{S09a}
                           }
\startdata
$\kappa_{v1}$& 4.00$\times$10$^{-3}$ & 4.71$\times$10$^{-3}$  &1.67$\times$10$^{-4}$ - 1.32$\times$10$^{-1}$ &0.475 &- &-\\
$\kappa_{v2}$ & 4.00$\times$10$^{-3}$ & 4.71$\times$10$^{-3}$  &1.67$\times$10$^{-4}$ - 1.32$\times$10$^{-1}$ &0.475 &- &-\\
$\kappa_{IR}$  & 3.00$\times$10$^{-2}$ &4.70$\times$10$^{-2}$ & 3.00$\times$10$^{-2}$ - 7.36$\times$10$^{-2}$&0.990 &- &-\\
$\alpha$  &0.5 & 0.5 &0.00 -1.00 &0.00 &-&-\\
$f_{H2O}$  & 4.00$\times$10$^{-4}$ & 1.19$\times$10$^{-4}$&5.29$\times$10$^{-5}$ - 2.67$\times$10$^{-4}$ &0.997 &$\sim10^{-4}$ &1$\times$10$^{-5}$ - 1$\times$10$^{-4}$ \\
$f_{CH4}$  &  1.00$\times$10$^{-6}$&9.78$\times$10$^{-9}$ & 9.79$\times$10$^{-15}$ - 9.77$\times$10$^{-3}$&0.00 & $\leq$6$\times$10$^{-6}$ &$\leq$1$\times$10$^{-7}$\\
$f_{CO}$  & 5.00$\times$10$^{-4}$ &1.15$\times$10$^{-2}$ &3.60$\times$10$^{-3}$ - 3.64$\times$10$^{-2}$ &0.993 &2$\times$10$^{-4}$ - 2$\times$10$^{-2}$ &1$\times$10$^{-4}$ - 3$\times$10$^{-4}$ \\
$f_{CO2}$  &  1.00$\times$10$^{-7}$& 3.37$\times$10$^{-3}$&1.69$\times$10$^{-3}$ - 6.72$\times$10$^{-3}$ &0.998 & 7$\times$10$^{-4}$&1$\times$10$^{-7}$ - 1$\times$10$^{-6}$\\
\enddata
\end{deluxetable*}
\end{center}



\section{Discussion \& Conclusions}

We demonstrate retrieval by inverse modeling of extrasolar planetary spectra. We first apply the technique to a synthetic model spectrum of a solar
metallicity $T \simeq 1200$ K hot Jupiter, and then to a previously published HST NICMOS spectrum of HD~189733b showing results that are consistent with
previous studies.  The approach herein is much more efficient that other methods such as a gridded parameter search, or Monte-Carlo techniques,
as it only requires $\sim 10^2$ forward model computations as opposed to millions. The formalism also allows robust estimation of the 
retrieval uncertainties.


We have also investigated the information theory aspects of the problem,  in order to assess the quality and usefulness of a spectral data set in constraining 
atmospheric properties. First, we discuss how the Jacobian matrix can be used to determine which spectral channels are most sensitive to 
chosen atmospheric parameters. Second, we show the use of the averaging kernel as a diagnostic tool to guide us to which parameters can be
usefully retrieved from the spectrum in question.    
Third, we calculated the number of available degrees of freedom and often found that, 
given the current limited observational capabilities, the number of retrievable parameters was less than the number of parameters in our forward model. 
Fourth, using simple expressions for the degrees of freedom and information content, we showed semi-quantitatively how $S/N$ and $R$ effect our knowledge 
of the atmospheric state.  These tools can be particularly useful in aiding the design of future instruments such that they can be optimized for observations
of transiting exoplanets.

A recent paper (Lee et al. 2011) using the optimal estimation approach as applied to HD~189733b, was published while this article 
was in preparation. The details of the methodology in that paper are somewhat different from ours, i.e. in the parameterization of the 
atmospheric models and in the use of the correlated-K
opacities (we use line-by-line radiative transfer). In addition, Lee et al. use multi-band (i.e. from various instruments 
inclusive of HST NICMOS, Spitzer IRAC, IRS and MIPS), multi-epoch measurements of HD~189733b as a representative snapshot of
the planetary dayside. We restrict our retrieval to a single epoch, 13 spectral-channel NICMOS observation spanning less than
one octave of total spectral coverage between 1.45-2.5 microns. Our retrievals agree for the most part with those of Lee et al., in that
 H$_2$O and CO$_2$ are retrieved with confidence but neither retrieval can say much about the abundance of methane (a trace species in
 HD~189733b).  One clear discrepancy is that we are able to retrieve CO where as they cannot. Also, Lee et al. do not discuss the information 
 content aspects of the atmospheric retrieval formulation presented in both of these papers. 
 
In follow on investigations, we plan to use the information content analyses to study aspects of combining Spitzer broadband photometry with prior
notions about the atmospheric state to constrain atmospheric properties such as CH$_4$/CO and C/O ratios. A powerful use of these methods is in
optimizing the design of instruments that could be flown in NASA's {\it FINESSE} and ESA's {\it Exoplanet Characterization Observatory}, or in studying the 
potential of already designed instruments such as {\it JWST's NIRCAM} that offer various observing modes, bandpasses and spectral resolving power.  




\section{Acknowledgements}		
We would like to thank Zhan Su, Aaron Wolf, Konstantin Batygin, Alejandro Soto, Run-Lie Shia, Leigh Fletcher, Kuai Le, Heather Knutson, Mimi Gerstell, Linda Brown, and the Yuk Yung group for reading the article and many useful discussions.   M. Line is supported by the JPL Graduate Fellowship funded by the JPL Research and Technology Development Program.   XZ and YLY are supported by a grant from the PATM program of NASA to the California Institute of Technology. P. Chen \& G. Vasisht are supported by the JPL Research \& Technology Development Program, and contributions herein were supported by the Jet Propulsion Laboratory, California Institute of Technology, under a contract with the National Aeronautics and Space Administration.


\begin{thebibliography}

\bibitem[Borysow (2002)]{2002A&A...390..779B} Borysow, A.\ 2002, \aap, 390, 779 
\bibitem[Borysow (2001)]{2001JQSRT...68..235} Borysow, A., J{\o}rgensen, U.G., Fu, Y.\ 2001, \jqsrt, 68, 235 
\bibitem[Chahine(1968)]{1968JOSA...58.1634C} Chahine, M.~T.\ 1968, JOSA (1917-1983), 58, 1634 
\bibitem[Fortney et al.(2010)]{2010ApJ...709.1396F} Fortney, J.~J., 
Shabram, M., Showman, A.~P., et al.\ 2010, \apj, 709, 1396 
\bibitem[Grillmair et al.(2007)]{2007ApJ...658L.115G} Grillmair, C.~J., 
Charbonneau, D., Burrows, A., et al.\ 2007, \apjl, 658, L115 
\bibitem[Grillmair et al.(2008)]{2008Natur.456..767G} Grillmair, C.~J., 
Burrows, A., Charbonneau, D., et al.\ 2008, \nat, 456, 767 
\bibitem[Guillot(2010)]{2010A&A...520A..27G} Guillot, T.\ 2010, \aap, 520, A27
\bibitem[Harrington et al.(2006)]{2006Sci...314..623H} Harrington, J., 
Hansen, B.~M., Luszcz, S.~H., et al.\ 2006, Science, 314, 623 
\bibitem[Harrington et al.(2007)]{2007Natur.447..691H} Harrington, J., 
Luszcz, S., Seager, S., Deming, D., 
\& Richardson, L.~J.\ 2007, \nat, 447, 691
\bibitem[Jacob (2007)]{} Jacob, D. \ 2007, Inverse modeling of atmospheric composition data, ARCNESS Winter School
\bibitem[J{\o}rgensen et 
al.(2000)]{2000A&A...361..283J} J{\o}rgensen, U.~G., Hammer, D., Borysow, A., \& Falkesgaard, J.\ 2000, \aap, 361, 283  
\bibitem[Knutson et al.(2007)]{2007Natur.447..183K} Knutson, H.~A.,Charbonneau, D., Allen, L.~E., et al.\ 2007, \nat, 447, 183 
 \bibitem[Knutson et al.(2008)]{2008ApJ...673..526K} Knutson, H.~A., 
Charbonneau, D., Allen, L.~E., Burrows, A., 
\& Megeath, S.~T.\ 2008, \apj, 673, 526 
\bibitem[Knutson et al.(2009)]{2009ApJ...690..822K} Knutson, H.~A., et al.\ 2009, \apj, 690, 822 
\bibitem[Kuai et al.(2010)]{2010JQSRT.111.1296K} Kuai, L., Natraj, V., 
Shia, R.-L., Miller, C., \& Yung, Y.~L.\ 2010, \jqsrt, 111, 1296 
\bibitem[Lee et al.(2011)]{2011arXiv1110.2934L} Lee, J.-M., Fletcher, 
L.~N., \& Irwin, P.~G.~J.\ 2011, arXiv:1110.2934 
\bibitem[Line et al.(2010)]{2010ApJ...717..496L} Line, M.~R., Liang, M.~C., 
\& Yung, Y.~L.\ 2010, \apj, 717, 496
\bibitem[Line et al.(2011)]{2011ApJ...738...32L} Line, M.~R., Vasisht, G., 
Chen, P., Angerhausen, D., \& Yung, Y.~L.\ 2011a, \apj, 738, 32 
\bibitem[Line et al.(2011)]{2011DPS} Line, M.~R.,Zhang, X., 
\& Yung, Y.~L.\ 2011b, EPSC-DPS Joint Meeting, 249
\bibitem[Madhusudhan 
\& Seager(2009)]{2009ApJ...707...24M} Madhusudhan, N., \& Seager, S.\ 2009, \apj, 707, 24 
\bibitem[Mandell et al.(2011)]{2011ApJ...728...18M} Mandell, A.~M., Drake 
Deming, L., Blake, G.~A., et al.\ 2011, \apj, 728, 18 
\bibitem[Moses et al.(2011)]{2011ApJ...737...15M} Moses, J.~I., Visscher, 
C., Fortney, J.~J., et al.\ 2011, \apj, 737, 15 
\bibitem[Nixon et al.(2007)]{2007Icar..188...47N} Nixon, C.~A., Achterberg, 
R.~K., Conrath, B.~J., et al.\ 2007, Icarus, 188, 47 
\bibitem[Press et al.(1995)]{1995} Press, W.H., Teukolksy, S., Vetterling, W.T., Fannery, B. \ 1995, Numerical Recipies: The art
of scientific computing, sec. ed., Camb. Univ. Press
\bibitem[Rodgers(1976)]{1976RvGSP..14..609R} Rodgers, C.~D.\ 1976, Reviews 
of Geophysics and Space Physics, 14, 609 
\bibitem[Rodgers (2000)]{} Rodgers, C.D., Inverse methods for atmospheric sounding, Theory and Practice,\ 2000
\bibitem[Rothman et al.(2009)]{2009JQSRT.110..533R} Rothman, L.~S., Gordon, 
I.~E., Barbe, A., et al.\ 2009, \jqsrt, 110, 533 
\bibitem[Rothman et al.(2010)]{2010JQSRT.111.2139R} Rothman, L.~S., Gordon, 
I.~E., Barber, R.~J., et al.\ 2010, \jqsrt, 111, 2139 
\bibitem[Shannon et al.(1962)]{2011DPS} Shannon, C.E., \& Weaver, W.,
 A Mathematical Theory of Communication, University of Illinois Press, 1962
\bibitem[Sharp 
\& Burrows(2007)]{2007ApJS..168..140S} Sharp, C.~M., \& Burrows, A.\ 2007, \apjs, 168, 140 
\bibitem[Sing et al.(2008)]{2008ApJ...686..667S} Sing, D.~K., Vidal-Madjar, 
A., Lecavelier des Etangs, A., et al.\ 2008, \apj, 686, 667 
\bibitem[Stevenson et al.(2010)]{2010Natur.464.1161S} Stevenson, K.~B., et al.\ 2010, \nat, 464, 1161
\bibitem[Swain et al.(2009a)]{2009ApJ...690L.114S} Swain, M.~R., Vasisht, 
G., Tinetti, G., et al.\ 2009a, \apjl, 690, L114 
\bibitem[Swain et al.(2009b)]{2009ApJ...704.1616S} Swain, M.~R., Tinetti, 
G., Vasisht, G., et al.\ 2009b, \apj, 704, 1616 
\bibitem[Swain et al.(2010)]{2010Natur.463..637S} Swain, M.~R., Deroo, P., 
Griffith, C.~A., et al.\ 2010, \nat, 463, 637 
\bibitem[Tinetti et al.(2007)]{2007Natur.448..169T} Tinetti, G., 
Vidal-Madjar, A., Liang, M.-C., et al.\ 2007, \nat, 448, 169 
\bibitem[Tinetti et al.(2010)]{2010ApJ...712L.139T} Tinetti, G., Deroo, P., 
Swain, M.~R., et al.\ 2010, \apjl, 712, L139 
\bibitem[Twomey et al.(1977)]{1977JAtS...34.1085T} Twomey, S., Herman, B., 
\& Rabinoff, R.\ 1977, JAS, 34, 1085



\end{thebibliography}
\end{document}